\documentclass[12pt]{article}
\usepackage{amsthm,amssymb,amsmath}
\usepackage[textwidth=18cm,textheight=22cm]{geometry}
\usepackage[english]{babel}
 1 
1

\newcommand{\C}{\ensuremath{\mathbb{C}}}

\newcommand{\p}{{\boldsymbol{p}}}

\newcommand{\cL}{{\cal L}}
\newcommand{\cP}{{\cal P}}

\newcommand{\bu}{{{\bf u}}}
\newcommand{\br}{{{\bf r}}}
\newcommand{\ba}{{{\bf a}}}

\newtheorem{pro}{Proposition}

\newcommand{\be}{\begin{equation}}
\newcommand{\ee}{\end{equation}}
\begin{document}

\title{\sc A Classification  of Integrable \\Quasiclassical Deformations\\ of Algebraic
Curves.
\thanks{Partially supported by  MEC
project FIS2005-00319 and by the grant COFIN 2004 "Sintesi" }}
\author{ B. Konopelchenko $^{1}$, L. Mart\'{\i}nez Alonso$^{2}$\\ and E. Medina$^{3}$
\\
\emph{ $^1$ Dipartimento di Fisica, Universit\'a di Lecce and
Sezione INFN}
\\ {\emph 73100 Lecce, Italy}\\
\emph{$^2$ Departamento de F\'{\i}sica Te\'{o}rica II, Universidad
Complutense}\\ \emph{E28040 Madrid, Spain}\\
\emph{$^3$ Departamento de Matem\'aticas, Universidad de C\'adiz}\\
\emph{E11510 Puerto Real, C\'adiz,Spain} }
\date{} \maketitle
\begin{abstract}
A previously introduced scheme for describing integrable deformations of
 of algebraic curves is completed. Lenard relations are used to characterize
and classify these deformations in terms of hydrodynamic type systems. A general solution
of the compatibility conditions for consistent deformations is
given and expressions for the solutions of the corresponding
Lenard relations are provided.
\end{abstract}

\vspace*{.5cm}

\begin{center}\begin{minipage}{12cm}

\end{minipage}
\end{center}
\newpage

\section{Introduction}

Algebraic curves find important applications in the theory of integrable systems \cite{1}-\cite{3}. They are particularly
relevant \cite{4}-\cite{7} in the study of the zero-dispersion limit of integrable systems and the analysis of Whitham equations. In  \cite{6}-\cite{7}
Krichever formulated a
general method  to characterize  dispersionless integrable systems underlying the deformations of algebraic curves in the Whitham averaging method.
 A different
scheme to determine integrable deformations of algebraic
curves $\mathcal{C}$ of the form
\begin{equation}\label{curve}
F(p,k):=p^N-\sum_{n=1}^{N}u_n(k)p^{N-n}=0.
\end{equation}
was introduced in \cite{8}-\cite{11}. Here the coefficients (\emph{potentials})
are assumed to be general polynomials in $k$.
Our previous work focused on curves of degrees $N=2$ and $3$  and  the aim of the present paper is to complete the analysis by  considering the
general case of algebraic curves of arbitrary degree $N$.

The method proposed in  \cite{8}-\cite{11} applies for finding  deformations  $\mathcal{C}(x,t)$ of \eqref{curve} such that the branches of the multiple-valued
function $p(k)=\left(p_1(k), \dots, p_N(k)\right)^T$
determined by \eqref{curve} obey
an equation of the form
\begin{equation}\label{2}
\partial_t p_i=\partial_x \Big(\sum_{r=1}^{N}a_r(k,u(k))p_i^{N-r}\Big),\quad a_r\in\C[k],
\end{equation}
where $a_r$ are functions of $k$ and $u(k)=(u_1(k),\ldots,u_N(k))$. As a consequence of \eqref{2} the potentials $u(k)$ satisfy an evolution equation of
hydrodynamic type and the problem  is to determine expressions for $a_r$  such that
\eqref{2} is consistent with the polynomial dependence of $u$
on the variable $k$. That is to say, if $(d_1,\ldots,d_N)$ are the degrees of the polynomials $(u_1(k),\ldots,u_N(k))$, then
 $\mbox{degree}(\partial_t u_n)\leq d_n $ must be satisfied for all $n$. At this point a Lenard relation allows us to formulate a sufficient condition
for the consistency of \eqref{2} in terms of a system of inequalities involving the degrees $d_n$ only. Thus we are led to the problem
of determining
the degrees satisfying the consistency condition (\emph{consistent degrees}) for each $N$.
In \cite{9} it was found that for $N=2$ the consistent degrees $(d_1,d_2)$ are characterized by the inequality $d_1\leq d_2+1$ . For $N=3$ there is only a finite set
 of consistent degrees  given by \cite{11}
\begin{equation}\label{N23}
\begin{array}{lll}
 &  & \begin{array}[t]{llllll} (0,0,1) & (0,1,0) & (0,1,1) & (0,1,2) & (1,0,0) & (1,0,1)\\
                                  (1,1,0) & (1,1,1) & (1,1,2) & (1,2,1) & (1,2,2) & (1,2,3). \end{array}
\end{array}\end{equation}
In the present work, we complete these results. Thus, it is first shown that for $N=4$ the set of consistent degrees is
\begin{equation}\label{N4}
\begin{array}{lll}
 &  & \begin{array}[t]{llll} (0,0,0,1) & (0,0,1,0) & (0,0,1,1) & (0,1,0,0)\\
                                (0,1,0,1) & (0,1,1,0) & (0,1,1,1) & (0,1,1,2),\end{array}
\end{array}\end{equation}
and then it is proved that for $N\geq5$ the consistent degrees
$(d_1,\ldots,d_N)$ are given by
\begin{equation}\label{Ngeq5}
d_i=0,\quad i=1,2,\dots,N-3,\qquad d_{N-2},\;d_{N-1},\;d_N\leq1.\end{equation}

We notice the fact that no compatible degrees $d_i\geq 2$
arise for $N\geq 5$, so that the degree $N=5$ represents a
threshold for a change in the properties of algebraic curves. This feature is reminiscent of
 the statement of the classical Abel
theorem \cite{15}.

By substituting the branches $p_i$ by their   Laurent series in $k$ into \eqref{2}, infinite series of
conservation laws follow. It means that the deformations of \eqref{curve} supplied by our method are integrable. In fact,
the corresponding hydrodynamic systems satisfied by the potentials $u_n(k)$  represent the quasiclassical (dispersionless) limits of the standard
integrable models arising from the compatibility between generalized (energy-dependent) spectral problems
\begin{equation}\label{esp}
\Big(\partial_x^N-\sum_{n=1}^{N}u_n(k,x)\partial_x^{N-n}\Big)\psi=0,
\end{equation}
and  equations of the form
\begin{equation}\label{esp2}
\partial_t \psi= \Big(\sum_{r=1}^{N}a_r(k,x,t)\partial_x^{N-r}\Big)\psi.
\end{equation}

The work is organized as follows.  We first outline our method in Section 2. Then Section 3 is devoted  to determine and classify the curves \eqref{curve} which admit  deformations consistent with the degrees of their potentials. Finally,
in Section 4 we characterize the hydrodynamic type systems which govern these deformations.

\section{Deformations of algebraic curves}

In order to write equation (\ref{2}) in terms of the potentials
$u_n$  we introduce the \emph{power sums}
\begin{equation}\label{ps1}
\mathcal{P}_s=\frac{1}{s}(p_1^s+\cdots+p_{N}^s),\quad s\geq 1.
\end{equation}
 One
can relate potentials and power sums through  Newton recurrence
formulas, the solution of which is given by Waring's formula
\cite{16}
\begin{equation}\label{ps2}
\mathcal{P}_s=\sum_{1\leq i\leq s}^{(s)} \,\frac{1}{i}(u_1+\cdots
+u_{N})^i,
\end{equation}
where the superscript $(s)$ in the summation symbol indicates that only the terms of weight $s$ are
retained, with the weights being defined as
\begin{equation}\label{weights}
weight[u_1^{\alpha_1}u_2^{\alpha_2}\cdots
u_N^{\alpha_N}]:=\sum_{j=1}^Nj\alpha_j.
\end{equation}
Using these variables, equation (\ref{2}) can be rewritten as \cite{10,11}
\begin{equation}\label{def2}
\partial_t\bu=J_0\ba,
\end{equation}
where
$$\everymath{\displaystyle}\begin{array}{l}
J_0=T^TV^T\partial_x\cdot V,\quad
\bu=\left(u_1, u_2, \dots u_N\right)^T,\quad
\ba=\left(a_N, a_{N-1}, \dots, a_1\right)^T,\\  \\
T:=\left(
\begin{array}{cccc}
1      &-u_{1}  &\cdots&-u_{N-1}\\
0      &1       &\cdots&-u_{N-2}\\
\vdots & \vdots &      & \vdots\\
0      &       0&\cdots&1
\end{array}
\right)\qquad
V:=\left(
\begin{array}{cccc}
1      &    p_1&\cdots&p_1^{N-1}\\
1      &    p_2&\cdots&p_2^{N-1}\\
\vdots &\vdots &      & \vdots \\
1      &    p_N&\cdots&p_N^{N-1}
\end{array}
\right).
\end{array}$$
The elements of $J_0$ can be easily written in terms of the power sums as
\begin{equation}\label{j}\everymath{\displaystyle}\begin{array}{llll}
(J_0)_{11}&=&N\partial_x,&\\  \\
(J_0)_{i1}&=&(i-1)\cP_{i-1}\partial_x-\sum_{l=2}^{i-1}u_{i-l}\cP_{l-1}\partial_x-Nu_{i-1}
\partial_x, & \mbox{if}\;\; i\neq1,\\  \\
(J_0)_{ij}&=&(i+j-2)\cP_{i+j-2}\partial_x+(j-1)\cP_{i+j-2,x}&  \\  \\
          & &-\sum_{k=1}^{i-1}u_{i-k}\left[(k+j-2)\cP_{k+j-2}\partial_x+(j-1)\cP_{k+j-2,x}\right],&
             \mbox{if}\;\;j\neq1.
\end{array}\end{equation}
The problem now is to determine expressions for $\ba$ (in
(\ref{def2})) depending on $k$ and $\bu$, such that the flow
\eqref{def2} is consistent with the polynomial dependence of $\bu$
on the variable $k$. That is to say, if $d_n:=\mbox{degree}(u_n)$
are the degrees of the coefficients $u_n$ as polynomials in $k$,
then
\[
\mbox{degree}(J_0 \ba)_n\leq d_n,\quad  n=1,\ldots N,
 \]
must be satisfied. The strategy \cite{9}-\cite{11} for finding consistent  deformations is to solve Lenard
type relations
\begin{equation}\label{lenard}
J_0\br=0,\quad \br:=(r_1,\ldots,r_N)^\top,\;\; r_i\in\C((k)),
\end{equation}
and take $\ba:=\br_+$, where $(\,\cdot\,)_+$ and $(\,\cdot\,)_-$
indicate the parts of non-negative and negative powers in $k$,
respectively. Now from the identity
\[
J_0\ba =J_0\br_+=-J_0\br_-,
\]
it is clear that  a sufficient condition for the consistency of \eqref{def2} is that
\begin{equation}\label{comp}
\max_{m=1,\dots,N}\{degree(J_0)_{nm}\}\leq d_n+1,\quad
n=1,\dots,N.
\end{equation}
This condition for
consistency only depends on the curve \eqref{curve} and
does not refer to the particular solution of the Lenard relation

In the subsequent discussion we will use
an important result concerning the branches $p_i(k)$: Let
$\C((\lambda))$ denote the field of Laurent series in $\lambda$ with
at most a finite number of terms with positive powers,
then we have \cite{13,14} :

\vspace{0.3cm}
\noindent
{\bf Newton Theorem}
\emph{ There exists a positive integer $l$ such that the $N$ branches
\begin{equation}\label{branches}
p_j(z):=\Big(p_j(k)\Big)\Big |_{k=z^l},
\end{equation}
are elements of  $\C((z))$. Furthermore, if $F(p,k)$ is irreducible
as a polynomial over the field $\C((k))$ then $l_0=N$ is the least
permissible $l$ and the branches $p_j(z)$ can be labelled so that
\[
p_j(z)=p_N(\epsilon^j z),\quad \epsilon:=\exp \left(\frac{2\pi
\imath}{N}\right).
\]
}

\noindent {\bf Notation convention} \emph{
 Henceforth, given an algebraic curve $\mathcal{C}$ we
will denote by $z$ the variable associated with the least positive
integer $l_0$  for which the substitution $k=z^{l_0}$ implies
$p_j\in\C((z)),\, \forall j$.  We refer to $l_0$ as the Newton
exponent of $\mathcal{C}$.}

It was proved in \cite{10}-\cite{11} that the solution of the Lenard relation $J_0\br=0$ is
given by
\begin{equation}\label{sollenard}
\br=T\nabla_{\bu}R,\quad R=\sum_{i=1}^Ng_i(z)p_i,\quad
\nabla_{\bu}R=\left(\frac{\partial R}{\partial u_1}, \dots, \frac{\partial R}{\partial u_N}\right)^T,
\end{equation}
with $g_i\in{\mathbb C}((z))$. The problem of choosing the
functions $g_i$ such that $R\in{\mathbb C}((k))$ (and consequently
$\br\in{\mathbb C}((k))$) was solved in \cite{11} by introducing
the element $\sigma_0$ of the Galois group of the curve
\begin{equation}\label{galois}
\sigma_0(p_j)(z):=p_j(\epsilon_0\,z),\quad
\epsilon_0:=\exp\left(\frac{2\pi \imath}{l_0}\right).\end{equation}
Thus it is clear that the requirement of $R\in{\mathbb C}((k))$ is
equivalent to the invariance of $R$ under $\sigma_0$ i.e.
\begin{equation}\label{invcon}
R(\epsilon_0\,z,\sigma_0\,\p)=R(z,\p).\end{equation} The scheme
now consists in using  the {\em Lagrange resolvents} \cite{15}
\begin{equation}\label{lag}
\mathcal{L}_i:=\sum_{j=1}^N (\epsilon^i)^j\,p_j,\quad i=1,2,\dots,N,
\end{equation}
to construct  functions $R$ satisfying \eqref{invcon} and such
that $R\in{\mathbb C}((k))$.

 The case $N=3$ was completely solved in \cite{11}. There arise twelve possible choices
(\ref{N23}) which are classified in terms of $\sigma_0$ and $l_0$
according to\vspace{3mm}
\begin{center}
{\bf Table 1:} Classification of (\ref{N23}) according to
$\sigma_0$ and $l_0$.
\end{center}
\begin{center}
\begin{tabular}{|c|c|c|}
\hline
$\sigma_0$  &  $l_0$   &  $(d_1,d_2,d_3)$\\
\hline
$\left(\begin{array}{lll}
p_1 & p_2 &  p_3  \\
p_2 & p_3 &  p_1
\end{array}\right)$
&  $3$  &  $\begin{array}{ll}(0,0,1)& (0,1,2)\end{array}$ \\ \hline
$\left(\begin{array}{lll}
p_1 & p_2 & p_3 \\
p_2 & p_1 & p_3
\end{array}\right)$
&  $2$  &
$\begin{array}{ll}
(0,1,0) & (0,1,1) \\   (1,0,0) & (1,1,2)
\end{array}$\\  \hline
$\left(\begin{array}{lll}
p_1 & p_2 &  p_3 \\
p_1 & p_2 &  p_3
\end{array}\right)$
&   $1$ &
$\begin{array}{ll}
(1,0,1) & (1,1,0) \\ (1,1,1) & (1,2,1) \\ (1,2,2) & (1,2,3)
\end{array}$\\
\hline
\end{tabular}
\end{center}

\vspace{3mm}

\noindent
and the invariant functions $R$ in (\ref{sollenard}) are given by
\begin{equation}\label{RN3}\begin{array}{lll}
l_0=3,&  &R=zf_1(z^3)\cL_1+z^2f_2(z^3)\cL_2+f_3(z^3)\cL_3,\\  \\
l_0=2,&  &R=f_1(z^2)(\cL_1+\cL_2)+zf_2(z^2)(\cL_1-\cL_2)+f_3(z^2)\cL_3\\  \\
l_0=1,&  &R=f_1(z)\cL_1+f_2(z)\cL_2+f_3(z)\cL_3,
\end{array}\end{equation}
with $f_1$, $f_2$ and $f_3$ being arbitrary analytic functions of $k$.

\section{Solutions of the consistency condition}

Let us first consider condition (\ref{comp}) for $N=4$. Taking
into account (\ref{j})
 we find that the elements of $J_0$  are given by
$$\begin{array}{l}
(J_0)_{11}=4\,\partial_x, \\  \\
(J_0)_{12}=u_1\,\partial_x+u_{1x},\\  \\
(J_0)_{13}=(u_1^2+2u_2)\partial_x+(u_1^2+2u_2)_x,   \\  \\
(J_0)_{14}=(u_1^3+3u_1\,u_2+3u_3)\partial_x+(u_1^3+3u_1\,u_2+3u_3)_x,
\\  \\
(J_0)_{21}=-3u_1\,\partial_x, \\   \\
(J_0)_{22}=2u_2\,\partial_x+u_{2x},\\  \\
(J_0)_{23}=(u_1\,u_2+3u_3)\,\partial_x+2(u_2\,u_{1x}+u_{3x}),
\\   \\
(J_0)_{24}=(u_1^2\,u_2+2u_2^2+u_1\,u_3+4u_4)\partial_x+
3(u_{4x}+u_2\,u_{2x}+u_2\,u_1\,u_{1x}+u_3\,u_{1x}),\\  \\
(J_0)_{31}=-2u_2\,\partial_x,  \\  \\
(J_0)_{32}=3u_3\,\partial_x+u_{3x},\\  \\
(J_0)_{33}=(4u_4+u_1\,u_3)\partial_x+2(u_{4x}+u_3\,u_{1x}),
\\  \\
(J_0)_{34}=(u_1\,u_4+2u_2\,u_3+u_1^2\,u_3)\partial_x+
3(u_4\,u_{1x}+u_3\,u_1\,u_{1x}+u_3u_{2x}),\\  \\
(J_0)_{41}=-u_3\,\partial_x,  \\  \\
(J_0)_{42}=4u_4\,\partial_x+u_{4x},\\  \\
(J_0)_{43}=u_1\,u_4\,\partial_x+2u_4\,u_{1x},\\  \\
(J_0)_{44}=(u_1^2\,u_4+2u_2\,u_4)\partial_x+3u_4(u_1\,u_{1x}+u_{2x}).
\end{array}$$
Thus, the compatibility condition (\ref{comp}) reduces to
$$\begin{array}{l}
d_1=0,\quad d_2\leq1,\quad d_3\leq 1,\\  \\
d_4\leq d_2+1,\quad d_4\leq d_3+1,
\end{array}$$
which leads to the proposition

\begin{pro}
For $N=4$ the degrees $(d_1,d_2,d_3, d_4)$ satisfying the
compatibility condition (\ref{comp}) are
\begin{equation}\label{cases4}\begin{array}{llll}
(0,0,0,1),  &  (0,0,1,0),  &  (0,0,1,1),  &  (0,1,0,0),\\  \\
(0,1,0,1),  &  (0,1,1,0),  &  (0,1,1,1),  &  (0,1,1,2).
\end{array}\end{equation}
\end{pro}

In order to derive our general result for  $N\geq5$, we start by
proving

\begin{pro}
For each $N\in{\mathbb N}$ $(N\geq 5)$ the degrees:
\begin{equation}\label{soldeg}
d_i=0,\quad i=1,2,\dots,N-3,\qquad d_{N-2},\;d_{N-1},\;d_N=0,1,\end{equation}
satisfy the compatibility condition (\ref{comp}).
\end{pro}

\begin{proof}

 We extend recursively the definition of the weights (\ref{weights}) by
\[
weight[(\partial_x^n
u_j)\,\mathrm{P}(\bu,\bu_x,\ldots)]=j+weight[\mathrm{P}(\bu,\bu_x,\ldots)],
\]
where $\mathrm{P}(\bu,\bu_x,\ldots)$ denotes any differential
polynomial in $\bu$. Taking into account (\ref{ps2}) and (\ref{j}),
we find that the elements of $J_0$ are weight homogeneous with
respect to the scaling:
$$(u_1,u_2,\dots,u_N)\;\rightarrow\;(\lambda u_1,\lambda^2 u_2,\dots,\lambda^N u_N),$$
and  their weights are given by
$$weight[(J_0)_{ik}]=i+k-2.$$
For the case $i+k\,<\,2N-2$ we have $weight[(J_0)_{ik}]<2N-4$ and,
as a consequence,  if the indexes $(i,k)$ satisfy $i+k\,<\,2N-2$
then $(J_0)_{ik}$ does not involve neither terms of the form
$u_{N-2}^{j+1}$, $u_{N-1}^{j+1}$, $u_N^{j+1}$, $u_{N-2}^ju_{N-1}^l$,
$u_{N-2}^ju_{N}^l$, $u_{N-1}^ju_{N}^l$, $j,l\geq 1$ nor similar
terms containing derivatives. Thus,
\begin{equation}\label{gen}\begin{array}{lll}
degree[(J_0)_{ik}]\leq&\max&\{[d_1,\dots,d_{N-3}],d_{N-2}+[d_1,\dots,d_{N-3}],\\
                      &    &\;d_{N-1}+[d_1,\dots,d_{N-3}],d_N+[d_1,\dots,d_{N-3}]\},
\end{array}\end{equation}
where  $[d_1,\dots,d_{N-3}]$ stands for degrees of terms appearing
in $(J_0)_{ik}$ which are linear combination of
$d_1$,...,$d_{N-3}$ with entire coefficients.

Now we examine the remaining elements  $(J_0)_{ik}$, i.e.
$$(i,k)\in\{(N-2,N),(N-1,N-1),(N-1,N),(N,N-2),(N,N-1),(N,N)\}.$$

\begin{itemize}

\item  $weight[(J_0)_{N-2,N}]=2N-4$, so that $(J_0)_{N-2,N}$ may contain terms of the form
$u_{N-2}^2$, $u_{N-2}u_{N-2,x}$ and we have
\begin{equation}\label{c1}\begin{array}{l}
degree[(J_0)_{N-2,N}]\leq\\  \\
\qquad \max\{[d_1,\dots,d_{N-3}],d_{N-2}+[d_1,\dots,d_{N-3}],\; d_{N-1}+[d_1,\dots,d_{N-3}],\\  \\
\qquad \qquad d_N+[d_1,\dots,d_{N-3}],\;2d_{N-2}\}.
\end{array}\end{equation}

\item $weight[(J_0)_{N-1,N-1}]=2N-4$. This weight allows the presence of terms
such as $u_{N-2}^2\partial_x$ and  $u_{N-2}u_{N-2,x}$, which arise
multiplied by the coefficients:

$$\begin{array}{l}
\mbox{coeff}[(2N-4)\cP_{2N-4}\partial_x,\;u_{N-2}^2\partial_x]=N-2,\\  \\
\mbox{coeff}[u_{N-k-1}(N+k-3)\cP_{N+k-3}\partial_x,\;u_{N-2}^2\partial_x]=
\left\{\begin{array}{lll}
N-2 & \mbox{if} & k=1,\\
0      & \mbox{if} & k\neq1,
\end{array}\right.\end{array}$$
$$\Rightarrow \mbox{coeff}[(J_0)_{N-1\,N-1},\;u_{N-2}^2\partial_x]=0.$$

$$\begin{array}{l}
\mbox{coeff}[(N-2)\cP_{2N-4,x},\;u_{N-2}u_{N-2\,x}]=N-2,\\  \\
\mbox{coeff}[u_{N-k-1}(N-2)\cP_{N+k-3,x},\;u_{N-2}u_{N-2\,x}]=
\left\{\begin{array}{lll}
N-2    & \mbox{if} & k=1,\\
0      & \mbox{if} & k\neq1,
\end{array}\right.\end{array}$$
$$\Rightarrow \mbox{coeff}[(J_0)_{N-1\,N-1},\;u_{N-2}u_{N-2\,x}]=0.$$
Thus, $(J_0)_{N-1\,N-1}$ does not contain terms in $u_{N-2}^2$, $u_{N-2}u_{N-2\,x}$
and consequently
\begin{equation}\label{c2}\begin{array}{l}
degree[(J_0)_{N-2,N}]\leq\\  \\
\qquad \max\{[d_1,\dots,d_{N-3}],\;d_{N-2}+[d_1,\dots,d_{N-3}],\;d_{N-1}+[d_1,\dots,d_{N-3}],\\  \\
\qquad \qquad d_N+[d_1,\dots,d_{N-3}]\}.
\end{array}\end{equation}

\item $weight[(J_0)_{N-1,N}]=2N-3$. Terms of the form $u_{N-2}^2u_1$, $u_{N-2}u_{N-1}$, or similar
terms containing derivatives may arise. A direct computation,
similar to the one in the previous case proves that there are no
terms $u_{N-2}^2u_1$, $u_{N-2}^2u_{1,x}$, $u_{N-2}u_{N-2,x}u_1$ in
$(J_0)_{N-1,N-1}$. Then we have that
\begin{equation}\label{c3}\begin{array}{l}
degree[(J_0)_{N-1,N}]\leq\\  \\
\qquad \max\{[d_1,\dots,d_{N-3}],\;d_{N-2}+[d_1,\dots,d_{N-3}],\;d_{N-1}+[d_1,\dots,d_{N-3}],\\  \\
\qquad \qquad d_N+[d_1,\dots,d_{N-3}],\;d_{N-2}+d_{N-1}\}.
\end{array}\end{equation}

\item  $weight[(J_0)_{N,N-2}]=2N-4$. A direct computation shows that
there are no terms $u_{N-2}^2$, $u_{N-2}u_{N-2,x}$ in
$(J_0)_{N,N-2}$, so that
\begin{equation}\label{c4}\begin{array}{l}
degree[(J_0)_{N,N-2}]\leq\\  \\
\qquad \max\{[d_1,\dots,d_{N-3}],\;d_{N-2}+[d_1,\dots,d_{N-3}],\;d_{N-1}+[d_1,\dots,d_{N-3}],\\  \\
\qquad \qquad d_N+[d_1,\dots,d_{N-3}]\}.
\end{array}\end{equation}

\item  $weight[(J_0)_{N,N-1}]=2N-3$. One can see that $(J_0)_{N,N-1}$
has  no terms $u_{N-2}^2u_1$, $u_{N-2}u_{N-1}$ or similar
terms containing derivatives. Consequently
\begin{equation}\label{c5}\begin{array}{l}
degree[(J_0)_{N,N-2}]\leq\\  \\
\qquad \max\{[d_1,\dots,d_{N-3}],\;d_{N-2}+[d_1,\dots,d_{N-3}],\;d_{N-1}+[d_1,\dots,d_{N-3}],\\  \\
\qquad \qquad d_N+[d_1,\dots,d_{N-3}]\}.
\end{array}\end{equation}

\item  $weight[(J_0)_{NN}]=2N-2$. This element may involve terms $u_{N-2}u_N$, $u_{N-2\,x}u_N$ or
$u_{N-2}u_{N\,x}$. On the other hand, it can be checked, as in the
previous cases, that terms  $u_{N-2}^2u_2$, $u_{N-2}^2u_1^2$,
$u_{N-2}u_{N-1}u_1$, $u_{N-1}^2$ or similar ones containing
derivatives cannot arise. Consequently
\begin{equation}\label{c6}\begin{array}{l}
degree[(J_0)_{N,N-2}]\leq\\  \\
\qquad \max\{[d_1,\dots,d_{N-3}],\;d_{N-2}+[d_1,\dots,d_{N-3}],\;d_{N-1}+[d_1,\dots,d_{N-3}],\\  \\
\qquad \qquad d_N+[d_1,\dots,d_{N-3}],\;d_{N-2}+d_N\}.
\end{array}\end{equation}

\end{itemize}

In summary, by taking into account (\ref{gen})-(\ref{c6}), we
conclude that (\ref{comp}) is satisfied provided that
\begin{equation}\label{ineq}\begin{array}{llll}
[d_1,\dots,d_{N-3}]\,\leq\, 1,      &   &   & 2d_{N-2}\,\leq\,d_{N-2}+1,\\  \\
d_{N-2}+[d_1,\dots,d_{N-3}]\,\leq\,1,&   &   & d_{N-2}+d_{N-1}\,\leq\,d_{N-1}+1,\\  \\
d_{N-1}+[d_1,\dots,d_{N-3}]\,\leq\,1,&   &   & d_{N-2}+d_{N}\,\leq\,d_{N}+1.\\  \\
d_{N}+[d_1,\dots,d_{N-3}]\,\leq\,1,&   &   &
\end{array}\end{equation}
Thus, any choice of the degrees verifying
$$d_i=0,\quad i=1,2,\dots,N-3,\qquad d_{N-2},d_{N-1},d_N\leq 1$$
satisfies (\ref{ineq}) and in consequence it verifies
(\ref{comp}).

\end{proof}

\vspace{3mm}

We next show that (\ref{soldeg}) constitutes the complete set of
degrees satisfying (\ref{comp}).

\begin{pro}
For each $N\in{\mathbb N}$ $(N\geq5)$  the compatibility condition
(\ref{comp}) implies
$$d_i=0,\quad i=1,2,\dots,N-3,\qquad d_{N-2},\;d_{N-1},\;d_N\leq1.$$
\end{pro}

\begin{proof}

 The cases $N$ even or odd must be considered separately. Suppose first that $N=2M$ with $M\in{\mathbb N}$ $(M\geq3)$. From
(\ref{j}) we have that
$$(J_0)_{1\,2M}=(2M-1)\cP_{2M-1}\partial_x+(2M-1)\cP_{2M-1,x}.$$
Thus, it is clear that $(J_0)_{1\,2M}$ contains terms in
$$u_1^{2M-1}\partial_x,\quad u_j^2u_1^{2M-2j-1}\partial_x,\quad
j=2,\dots,M-1,$$ $$ u_{2M-1}\partial_x,\quad
u_{2M-2}u_1\partial_x,$$ and consequently, the condition
(\ref{comp}) with $n=1$ implies that
$$\begin{array}{l}
(2M-1)d_1\leq d_1+1,\quad 2d_j+(2M-2j-1)d_1\leq d_1+1,\quad j=2,\dots,M-1,\\  \\
d_{2M-1}\leq d_1+1,\quad d_{2M-2}+d_1\leq d_1+1,
\end{array}$$
or equivalently
\begin{equation}\label{even1}
d_j=0,\quad j=1,2,\dots,M-1,\qquad d_{2M-2},d_{2M-1}\leq 1.
\end{equation}
By taking now $i=2l$, $j=2M$ $(l<M)$ in (\ref{j}) we have that
$$\everymath{\displaystyle}\begin{array}{lll}
(J_0)_{2l\,2M}&=&2(l+M-1)\cP_{2(l+M-1)}\partial_x+(2M-1)\cP_{2(l+M-1),x}\\  \\
              & &-\sum_{k=1}^{2l-1}u_{2l-k}\left[(k+2M-2)\cP_{k+2M-2}\partial_x+(2M-1)\cP_{k+2M-2,x}\right].
\end{array}$$
Then, we have that $(J_0)_{2\,2M}$ contains a term
$u_{2M}\partial_x$  so that
$$d_{2M}\leq d_2+1.$$
Since according to (\ref{even1}) $(M\geq3)$ $d_2=0$, we have that
\begin{equation}\label{even3}
d_{2M}\leq 1.
\end{equation}
On the other hand, we also see that $(J_0)_{2l\,2M}$ contains a
term  $u_{l+M-1}^2\partial_x$. Hence, the condition (\ref{comp})
with $n=2l$ implies
\begin{equation}\label{even2}
2d_{l+M-1}\leq d_{2l}+1,\quad \mbox{for each}\quad l<M.
\end{equation}

Now from (\ref{even2}) we  deduce:
\begin{itemize}
\item By setting  $l=1$ in (\ref{even2}),  we get $2d_M\leq d_2+1$, but $d_2=0$ so that $d_M=0$. Thus,
$$M\geq3\;\Rightarrow\; d_j=0,\quad j=1,2,\dots,M.$$
\item Suppose that $M\geq4$, and put $l=2$ into (\ref{even2}), then we have that $2d_{M+1}\leq d_4+1$.
But under our hypothesis $d_4=0$, so that
$$M\geq4\;\Rightarrow d_j=0,\quad j=1,2,\dots,M+1.$$
\item Suppose that $M\geq5$, and put $l=3$ into (\ref{even2}), then  $2d_{M+2}\leq d_6+1$.
Again, under our actual hypothesis $d_6=0$, we have that
$$M\geq5\;\Rightarrow\; d_j=0,\quad j=1,2,\dots,M+2.$$
\end{itemize}
Let us now use induction to prove
\begin{equation}\label{evenhyp}
M\geq k+3\;\Rightarrow\; d_j=0,\quad j=1,2,\dots,M+k.
\end{equation}
We have already proved (\ref{evenhyp}) for $k=1,2$. Assume that it
holds for $k\leq k_0-1$ and let us check it for $k=k_0$:

Take $M\geq k_0+3$, and  put $l=k_0+1$ in (\ref{even2}), then we
have that
$$2d_{M+k_0}\leq d_{2k_0+2}+1.$$
As $2k_0+2\leq M+k_0-1$ it follows that $d_{2k_0+2}=0$, so that
$d_{M+k_0}=0$ which proves (\ref{evenhyp}).

 Finally, for a given $M$,
 take $k=M-3$, then
$$d_j=0,\quad j=1,2,\dots, 2M-3.$$
Hence, by taking  (\ref{even1}) and (\ref{even3}) into account, we
have proved that (\ref{comp}) implies
$$d_j=0,\quad j=1,2,\dots,2M-3,\qquad d_{2M-2},\,d_{2M-1},\,d_{2M}\leq 1.$$

\vspace{3mm}

We consider now the case $N=2M+1$ with $M\in{\mathbb N}$
$(M\geq2)$. From (\ref{j})
$$(J_0)_{1\,2M+1}=2M\cP_{2M}\partial_x+2M\cP_{2M,x}.$$
Consequently $(J_0)_{1\,2M+1}$ contains terms in
$$u_1^{2M}\partial_x,\quad u_j^2u_1^{2M-2j}\partial_x,\quad j=2,\dots,M,\quad
u_{2M}\partial_x,\quad u_{2M-1}u_1\partial_x,$$
and the condition (\ref{comp}) with $n=1$ implies that
$$\begin{array}{l}
2Md_1\leq d_1+1,\quad 2d_j+(2M-2j)d_1\leq d_1+1,\quad j=2,\dots,M,\\  \\
d_{2M}\leq d_1+1,\quad d_{2M-1}+d_1\leq d_1+1,
\end{array}$$
or equivalently
\begin{equation}\label{odd1}
d_j=0,\quad j=1,2,\dots,M,\qquad d_{2M-1},d_{2M}\leq 1.
\end{equation}
On the other hand, by setting $i=2l+1$, $j=2M+1$ $(l<M)$ in
(\ref{j}) we have that
$$\everymath{\displaystyle}\begin{array}{lll}
(J_0)_{2l+1\,2M+1}&=&2(l+M)\cP_{2(l+M)}\partial_x+2M\cP_{2(l+M),x}\\  \\
              & &-\sum_{k=1}^{2l}u_{2l+1-k}\left[(k+2M-1)\cP_{k+2M-1}\partial_x+2M\cP_{k+2M-1,x}\right].
\end{array}$$
Thus, $(J_0)_{2l+1\,2M+1}$ contains the term
$u_{M+l}^2\partial_x$, so that the condition (\ref{comp}) with
$n=2l+1$ implies
\begin{equation}\label{odd2}
2d_{M+l}\leq d_{2l+1}+1.
\end{equation}
By putting  $l=1,2,3$ in (\ref{odd2})it follows
\begin{itemize}
\item For $l=1$ we have that $2d_{M+1}\leq d_3+1$. Thus,
$$M\geq3\;\Rightarrow\; d_j=0,\quad j=1,2,\dots,M+1.$$
\item For $l=2$ it follows that $2d_{M+2}\leq d_5+1$. Consequently
$$M\geq4\;\Rightarrow d_j=0,\quad j=1,2,\dots,M+2.$$
\item For $l=3$ the inequality (\ref{odd2}) reads $2d_{M+3}\leq d_7+1$ so that
$$M\geq5\;\Rightarrow\; d_j=0,\quad j=1,2,\dots,M+3.$$
\end{itemize}

Let us now use induction to show that
\begin{equation}\label{oddhyp}
M\geq k+2\;\Rightarrow\; d_j=0,\quad j=1,2,\dots,M+k.
\end{equation}
We have proved (\ref{oddhyp}) for $k=1,2,3$. Suppose that it holds
for $k\leq k_0-1$ and let us check it for $k=k_0$.  Take $M\geq
k_0+2$ and $l=k_0$ in (\ref{odd2}), we find
$$2d_{M+k_0}\leq d_{2k_0+1}+1.$$
But $2k_0+1\leq M+k_0-1$, then $d_{2k_0+1}=0$, $d_{M+k_0}=0$ and
(\ref{oddhyp}) follows. Thus, for a given $M$, if we take $k=M-2$
we have that
\begin{equation}\label{odd4}
d_j=0,\quad j=1,2,\dots, 2M-2.
\end{equation}

Finally, from the expression
$$\everymath{\displaystyle}\begin{array}{lll}
(J_0)_{2\,2M+1}&=&(2M+1)\cP_{2M+1}\partial_x+2M\cP_{2M+1,x}\\  \\
               & &-u_1\left[2M\cP_{2M}\partial_x+2M\cP_{2M,x}\right],
\end{array}$$
we have that (\ref{comp}) implies  $d_{2M+1}\leq d_2+1$, and
consequently $d_{2M+1}\leq1$. This fact, together with
(\ref{odd1}) and (\ref{odd4}) lead us to
$$d_j=0,\quad j=1,2,\dots,2M-2,\qquad d_{2M-1},\,d_{2M},\,d_{2M+1}\leq 1.$$

\end{proof}

From propositions 2 and 3 it follows that

\vspace{0.3cm}
\noindent
{\bf Theorem}

\emph{For each  $N\in{\mathbb N}$ $(N\geq5)$ the degrees
$(d_1,\ldots,d_N)$ satisfy the compatibility condition (\ref{comp})
if and only if
\begin{equation}\label{Ngeq5}
d_i=0,\quad i=1,2,\dots,N-3,\qquad d_{N-2},\;d_{N-1},\;d_N\leq1.\end{equation}
}

\section{Hierarchies of consistent deformations}

Our next task is to classify all the compatible cases in terms of
the corresponding  Newton exponent and the element $\sigma_0$
(\ref{galois}) of the Galois group of the curve.

We start by considering the case $N\geq5$. In order to find $l_0$
and $\sigma_0$ for each one of the seven nontrivial choices (\ref{Ngeq5}), we
study the asymptotic behavior of the $N$ branches $p_i$,
$i=1,2,\dots,N$ as $k\rightarrow\infty$.  By writing the
potentials as $$u_n=\displaystyle\sum_{j=0}^{d_n}u_{n\,j}k^j$$ we
have:

\begin{itemize}

\item $(0,\dots,0,0,0,1)$. In this case (\ref{curve}) can be written
as
$$k=\frac{1}{u_{N\,1}}\left(p^N-\sum_{l=1}^Nu_{l\,0}p^{N-l}\right),$$
so that
$$p_j^N \sim u_{N1}\,k \quad  \mbox{as} \quad k\rightarrow\infty,\quad
j=1,2,\dots,N.$$
Consequently, $p_j\in{\mathbb C}((k^{\frac{1}{N}}))$,
$j=1,2,\dots, N$ and
$$l_0=N, \qquad
\sigma_0=\left(\begin{array}{lllll}
p_1 & p_2 & \cdots & p_{N-1} & p_N \\
p_2 & p_3 & \cdots & p_N     & p_1
\end{array}\right).$$

\item $(0,\dots,0,0,1,0)$. Now, (\ref{curve}) takes the form
$$k=\frac{1}{u_{N-1\,1}}\left(p^{N-1}-\sum_{l=1}^Nu_{l\,0}p^{N-l-1}-
\frac{u_{N\,0}}{p}\right).$$
Thus, the roots satisfy
$$\everymath{\displaystyle}\begin{array}{l}
p_j^{N-1} \sim u_{N-1\,1}\,k \quad  \mbox{as} \quad k\rightarrow\infty,\quad
j=1,2,\dots,N-1,\\  \\
p_N\sim\,-\frac{u_{N\,0}}{u_{N-1\,1}}\frac{1}{k} \quad  \mbox{as} \quad
k\rightarrow\infty,
\end{array}$$
and we find
$$l_0=N-1, \qquad
\sigma_0=\left(\begin{array}{lllll}
p_1 & p_2 & \cdots & p_{N-1} & p_N \\
p_2 & p_3 & \cdots & p_1     & p_N
\end{array}\right).$$

\item $(0,\dots,0,0,1,1)$. From (\ref{curve}) we can write
$$k=\sum_{j=0}^{N-1}c_jp^j+\frac{c_{-1}}{u_{N-1\,1}p+u_{N\,1}},$$
for certain coefficients $c_j$, $j=-1,0,1,\dots,N-1$. Hence
$$\everymath{\displaystyle}\begin{array}{l}
p_j^{N-1} \sim \frac{1}{c_{N-1}}\,k \quad  \mbox{as} \quad k\rightarrow\infty,\quad
j=1,2,\dots,N-1,\\  \\
p_N\sim -\frac{u_{N\,1}}{u_{N-1\,1}}+\frac{c_{-1}}{u_{N-1\,1}}\frac{1}{k} \quad  \mbox{as} \quad
k\rightarrow\infty,
\end{array}$$
so that
$$l_0=N-1, \qquad
\sigma_0=\left(\begin{array}{lllll}
p_1 & p_2 & \cdots & p_{N-1} & p_N \\
p_2 & p_3 & \cdots & p_1     & p_N
\end{array}\right).$$

\item $(0,\dots,0,1,0,0)$. The equation (\ref{curve}) of the curve
implies
$$k=\frac{1}{u_{N-2\,1}}\left(p^{N-2}-\sum_{l=1}^{N-2}u_{l\,0}p^{N-l-2}+
\frac{u_{N-1\,0}}{p}+\frac{u_{N\,0}}{p^2}\right).$$
Then,
$$\everymath{\displaystyle}\begin{array}{l}
p_j^{N-2} \sim u_{N-2\,1}\,k \quad  \mbox{as} \quad k\rightarrow\infty,\quad
j=1,2,\dots,N-2,\\  \\
p_j^2\sim\frac{u_{N\,0}}{u_{N-2\,1}}\frac{1}{k} \quad  \mbox{as}
\quad k\rightarrow\infty,\quad j=N-1,N.
\end{array}$$
Thus, the corresponding Galois group element is given by
$$\sigma_0=\left(\begin{array}{llllll}
p_1 &  p_2 & \cdots & p_{N-2} & p_{N-1} & p_N \\
p_2 &  p_3 & \cdots & p_1     & p_N     & p_{N-1}
\end{array}\right),$$
and the Newton exponent is
$$l_0=\left\{\begin{array}{lll}
N-2 & \mbox{if} & N\quad \mbox{is even,}\\  \\
2(N-2) & \mbox{if} & N\quad \mbox{is odd.}
\end{array}\right.$$

\item $(0,\dots,0,1,1,0)$. From (\ref{curve}) we have
$$k=\sum_{j=0}^{N-2}c_jp^j+\frac{d_{1}}{p-b_1}+\frac{d_{2}}{p},$$
for certain coefficients $c_j$, $j=0,1,\dots,N-2$, $b_1$ and
$d_k$, $k=1,2$. The branches satisfy
$$\everymath{\displaystyle}\begin{array}{l}
p_j^{N-2} \sim \frac{1}{c_{N-2}}\,k \quad  \mbox{as} \quad k\rightarrow\infty,\quad
j=1,2,\dots,N-2,\\  \\
p_{N-1}\sim b_1+\frac{d_1}{k} \quad  \mbox{as} \quad
k\rightarrow\infty,\\  \\
p_{N}\sim \frac{d_2}{k} \quad  \mbox{as} \quad
k\rightarrow\infty,
\end{array}$$
so that
$$l_0=N-2, \qquad
\sigma_0=\left(\begin{array}{llllll}
p_1 & p_2 & \cdots & p_{N-2}&  p_{N-1} & p_N \\
p_2 & p_3 & \cdots & p_1    &  p_{N-1} & p_N
\end{array}\right).$$

\item $(0,\dots,0,1,0,1)$ and $(0,\dots,0,1,1,1)$. In these cases
(\ref{curve}) implies
$$k=\sum_{j=0}^{N-2}c_jp^j+\frac{d_{1}}{p-b_1}+\frac{d_{2}}{p-b_2},$$
for certain coefficients $c_j$,  $b_k$, $d_k$, $j=0,1,\dots,N-2$;
$k=1,2$. Therefore
$$\everymath{\displaystyle}\begin{array}{l}
p_j^{N-2} \sim \frac{1}{c_{N-2}}\,k \quad  \mbox{as} \quad k\rightarrow\infty,\quad
j=1,2,\dots,N-2,\\  \\
p_{N-1}\sim b_1+\frac{d_1}{k} \quad  \mbox{as} \quad
k\rightarrow\infty,\\  \\
p_{N}\sim b_2+\frac{d_2}{k} \quad  \mbox{as} \quad
k\rightarrow\infty,
\end{array}$$
so that
$$l_0=N-2, \qquad
\sigma_0=\left(\begin{array}{llllll}
p_1 & p_2 & \cdots & p_{N-2}&  p_{N-1} & p_N \\
p_2 & p_3 & \cdots & p_1    &  p_{N-1} & p_N
\end{array}\right).$$

\end{itemize}

\newpage
These results  are summarized in the following table

\vspace{5mm}

\begin{center}
{\bf Table 2:} Classification of (\ref{Ngeq5}) according to
$\sigma_0$ and $l_0$.
\end{center}
\begin{center}
\begin{tabular}{|c|c|c|}
\hline
$\sigma_0$  &  $l_0$   &  $(d_1,\dots, d_N)$\\
\hline
$\left(\begin{array}{lllll}
p_1 & p_2 & \dots &  p_{N-1} & p_N \\
p_2 & p_3 & \dots &  p_N    &  p_1
\end{array}\right)$
&  $N$  &  $\begin{array}{l}(0,\dots, 0,0,0,1)\end{array}$ \\ \hline
$\left(\begin{array}{lllll}
p_1 & p_2 & \dots & p_{N-1} & p_N \\
p_2 & p_3 & \dots & p_1     & p_N
\end{array}\right)$
&  $N-1$  &
$\begin{array}{l}
(0,\dots, 0,0,1,0) \\   (0,\dots, 0,0,1,1)
\end{array}$\\  \hline
$\left(\!\!\!\begin{array}{llllll}
p_1 &  \dots & p_{N-2}&  p_{N-1} & p_N \\
p_2 &  \dots & p_1    &  p_{N-1} & p_N
\end{array}\!\!\!\right)$
&   $N-2$ &
$\begin{array}{l}
(0,\dots, 0,1,1,0) \\  (0,\dots, 0,1,1,1) \\ (0,\dots, 0,1,0,1)
\end{array}$\\ \hline
$\left(\!\!\!\begin{array}{llllll}
p_1 &  \dots & p_{N-2}&  p_{N-1} & p_N \\
p_2 &  \dots & p_1    &  p_{N}   & p_{N-1}
\end{array}\!\!\!\right)$
&
\begin{tabular}{ll}
$N-2$ & if $N$ even \\
$2(N-2)$ & if $N$ odd
\end{tabular}
&  $(0,\dots, 0,1,0,0)$\\
\hline
\end{tabular}
\end{center}
\vspace{5mm}

We end this section by completing the previous table for $N=4$.
Only the special set of degrees $(0,1,1,2)$ remains to be analyzed.
The corresponding branches can be expanded as
$$p_i=a_{i\,1}k^{\frac{1}{2}}+a_{i\,0}+\frac{a_{i\,-1}}{k^{\frac{1}{2}}}+\cdots,\quad i=1,2,3,4,$$
where
$$\everymath{\displaystyle}\begin{array}{lll}
a_{i\,0}&=&\frac{{a_{i\,1}}^2\,u_{1\,0} +u_{3\,1}}{4\,a_{i\,1}^2 -2\,u_{2\,1}},\\  \\
a_{i\,-1}&=&\frac{1}{8\,a_{i\,1}\,{\left( 2\,a_{i\,1}^2 - u_{2\,1}\right) }^3}
\Big[a_{i\,1}^6\,\left( 6\,u_{1\,0}^2 + 16\,u_{2\,0} \right)\\  \\
  &  &  +a_{i\,1}^4\,\left( -5\,u_{1\,0}^2\,u_{2\,1} + 4\,u_{1\,0}\,u_{3\,1} +
       16\,\left( - u_{2\,0}\,u_{2\,1}  + u_{4\,1} \right)\right)
       \\  \\
  &   &- 2\,a_{i\,1}^2\,\left( -2\,u_{2\,0}\,u_{2\,1}^2 +
       3\,u_{1\,0}\,u_{2\,1}\,u_{3\,1} + u_{3\,1}^2 +
       8\,u_{2\,1}\,u_{4\,1} \right) \\  \\
  &  &  +u_{2\,1}\,\left( -u_{3\,1}^2 + 4\,u_{2\,1}\,u_{4\,1} \right)     \Big],\\  \\
\vdots &  & \vdots
\end{array}$$
and $a_{i\,1}$, $i=1,2,3,4$ are the solutions of the equation:
$$a_1^4-u_{2\,1}a_1^2-u_{4\,2}=0.$$
By labeling its solutions so that
$a_{2\,1}=-a_{1\,1}$, $a_{4\,1}=-a_{3\,1}$, we obtain
$$p_2(z)=p_1(-z),\quad p_4(z)=p_3(-z),\quad k=z^2.$$
Thus it follows that
$$l_0=2,\quad \sigma_0=\left(\everymath{\displaystyle}\begin{array}{llll}
p_1 & p_2 & p_3 & p_4 \\
p_2 & p_1 & p_4 & p_3
\end{array}\right).$$
Therefore, the table for $N=4$ is
\begin{center}
{\bf Table 3:} Classification of (\ref{N4}) according to
$\sigma_0$ and $l_0$.
\end{center}  \nopagebreak
\begin{center}
\begin{tabular}{|c|c|c|}
\hline
$\sigma_0$  &  $l_0$   &  $(d_1,d_2,d_3,d_4)$\\
\hline
$\left(\begin{array}{llll}
p_1 & p_2 &  p_3 & p_4 \\
p_2 & p_3 &  p_4 &  p_1
\end{array}\right)$
&  $4$  &  $\begin{array}{l}(0,0,0,1)\end{array}$ \\ \hline
$\left(\begin{array}{llll}
p_1 & p_2 & p_3 & p_4 \\
p_2 & p_3 & p_1 & p_4
\end{array}\right)$
&  $3$  &
$\begin{array}{l}
(0,0,1,0) \\   (0,0,1,1)
\end{array}$\\  \hline
$\left(\begin{array}{llll}
p_1 & p_2 &  p_3 & p_4 \\
p_2 & p_1 &  p_3 & p_4
\end{array}\right)$
&   $2$ &
$\begin{array}{l}
(0,1,1,0) \\  (0,1,1,1) \\ (0,1,0,1)
\end{array}$\\ \hline
$\left(\begin{array}{llll}
p_1 & p_2 & p_3 & p_4 \\
p_2 & p_1 & p_4 & p_3
\end{array}\right)$
& $2$ &
$\begin{array}{l}
(0,1,0,0) \\  (0,1,1,2)
\end{array}$\\
\hline
\end{tabular}
\end{center}

\vspace{0.3cm}

Let us now turn our attention to the problem  of obtaining the hierarchy of integrable deformations
(\ref{def2}). It is required to determine the function $R$ of the
form (\ref{sollenard}) satisfying the invariance condition
(\ref{invcon}).  In view of (\ref{invcon}) we discuss the different cases according to the corresponding element $\sigma_0$ of
the Galois group of the curve.

\begin{itemize}

\item $\sigma_0=\left(\everymath{\displaystyle}\begin{array}{lllll}
p_1  & p_2 & \dots & p_{N-1} & p_N\\ p_2 &  p_3 & \dots & p_N & p_1 \end{array}\right)$.

\end{itemize}

From the tables 1, 2 and 3 we have that $l_0=N$,
($\epsilon_0=\epsilon=e^{\frac{2\pi \imath}{N}}$). For $N\geq4$ the
only choice of degrees corresponding to $\sigma_0$ is
$(0,\dots,0,0,0,1)$. We look for
functions
$R_k=\sum_{j=1}^N\alpha_jp_j$ such that $ \sigma_0(R_k)=\epsilon_0^{N-k}R_k,\quad k=0,1,\dots,N-1.$
It is easy to check that
$$\sigma_0(R_k)=\alpha_Np_1+\sum_{j=2}^{N}\alpha_{j-1}p_j,$$
so that the condition $\sigma_0(R_k)=\epsilon_0^{N-k}R_k$ implies
that
$$\begin{array}{l}
\alpha_{j-1}=\epsilon_0^{N-k}\alpha_j,\qquad j=2,\dots N-1,N;\\  \\
\alpha_N=\epsilon_0^{N-k}\alpha_1.
\end{array}$$
This system admits the nontrival solutions
$$\alpha_j=\epsilon_0^{(N-k)(N-j)}\alpha_N=\epsilon_0^{j\,k}\alpha_N.$$
Thus the functions $R$ of the form (\ref{sollenard}) which satisfy
(\ref{invcon}) can be written as
\begin{equation}\label{RN}
R=\sum_{k=0}^{N-1}z^kf_k(z^N)\sum_{j=1}^N\epsilon_0^{j\,k}p_j,
\end{equation}
with $f_k\in{\mathbb C}((z^N))$, $k=0,1,\dots, N-1$. Taking into
account that $\epsilon_0=\epsilon$ and recalling (\ref{lag}), we
see that the functions $R$ can also be written in terms of the
Lagrange resolvents as
$$R=f_0(z^N)\cL_N+\sum_{k=1}^{N-1}z^kf_k(z^N)\cL_k,$$
which coincides with the first equation  for $N=3$ in (\ref{RN3}).

\begin{itemize}

\item $\sigma_0=\left(\everymath{\displaystyle}\begin{array}{lllll}
p_1  & \dots & p_{N-2}& p_{N-1} & p_N\\ p_2 &  \dots & p_{N-1} & p_1 & p_N \end{array}\right)$.
\end{itemize}

The corresponding Newton exponent is  $l_0=N-1$
($\epsilon_0=e^{\frac{2\pi \imath}{N-1}}$) and for $N\geq4$ the degrees of the
potentials  are $(0,\dots,0,0,1,0)$ and
$(0,\dots,0,0,1,1)$.
In this case we have that $\sigma_0(p_N)=p_N$, or equivalently
$p_N\in{\mathbb C}((k))$. Moreover, we need  $N-1$ additional functions $R$ verifying the
invariance condition (\ref{invcon}). Proceeding as in
the previous case we look for functions of the form
$$R_k=\sum_{j=1}^{N-1}\alpha_jp_j,\quad \mbox{such that}\quad
\sigma_0(R_k)=\epsilon_0^{N-1-k}R_k,\; k=0,1,\dots,N-2.$$ Since
the action of $\sigma_0$ on the function $R_k$ is given by
$$\sigma_0(R_k)=\alpha_Np_1+\sum_{j=2}^{N-1}\alpha_{j-1}p_j,$$
the condition $\sigma_0(R_k)=\epsilon_0^{N-1-k}R_k$ leads to
$$\begin{array}{l}
\alpha_{j-1}=\epsilon_0^{N-1-k}\alpha_j,\qquad j=N-1,N-2\dots,2\\  \\
\alpha_{N-1}=\epsilon_0^{N-1-k}\alpha_1,
\end{array}$$
so that $\alpha_j=\epsilon_0^{(N-1-k)(N-1-j)}\alpha_N=\epsilon_0^{j\,k}\alpha_N$, and
\begin{equation}\label{RNm1}
R=\sum_{k=0}^{N-2}z^kf_k(z^{N-1})\sum_{j=1}^{N-1}\epsilon_0^{j\,k}p_j+f_{N-1}(z^{N-1})p_N.
\end{equation}
%

\noindent
{\bf Example} For $N=4$
$$\begin{array}{lll}
R&=&f_0(z^3)(p_1+p_2+p_3)+zf_1(z^3)(e^{\frac{2\pi i}{3}}p_1+e^{\frac{4\pi i}{3}}p_2+p_3)\\  \\
 & &+z^2f_2(z^3)(e^{\frac{4\pi i}{3}}p_1+e^{\frac{2\pi i}{3}}p_2+p_3)+f_3(z^3)p_4.
\end{array}$$

\begin{itemize}
\item $\sigma_0=\left(\everymath{\displaystyle}\begin{array}{lllll}
p_1  & \dots & p_{N-2}& p_{N-1} & p_N\\ p_2 &  \dots & p_1 & p_{N-1} & p_N \end{array}\right)$.
\end{itemize}

In this case $\sigma_0$, $l_0=N-2$, ($\epsilon_0=e^{\frac{2\pi \imath}{N-2}}$). For
$N\geq4$ it corresponds to
the sets of degrees $(0,\dots,0,1,0,1)$, $(0,\dots,0,1,1,0)$ and $(0,\dots,0,1,1,1)$ .
Notice that $p_{N-1},\,p_N\in{\mathbb C}((k))$. Let us look for functions
$$R_k=\sum_{j=1}^{N-2}\alpha_jp_j,\quad \mbox{verifying}\quad
\sigma_0(R_k)=\epsilon_0^{N-2-k}R_k,\; k=0,1,\dots,N-3.$$
We find that
$$\begin{array}{l}
\alpha_{j-1}=\epsilon_0^{N-2-k}\alpha_j,\qquad j=N-2,N-3\dots,2\\  \\
\alpha_{N-2}=\epsilon_0^{N-2-k}\alpha_1,
\end{array}$$
then $\alpha_j=\epsilon_0^{(N-2-k)(N-2-j)}\alpha_{N-2}=\epsilon_0^{j\,k}\alpha_{N-2}$, and
\begin{equation}\label{RNm2a}
R=\sum_{k=0}^{N-3}z^kf_k(z^{N-2})\sum_{j=1}^{N-2}\epsilon_0^{j\,k}p_j+
f_{N-2}(z^{N-2})p_{N-1}+f_{N-1}(z^{N-2})p_N.
\end{equation}

\begin{itemize}
\item $\sigma_0=\left(\everymath{\displaystyle}\begin{array}{lllll}
p_1  & \dots & p_{N-2}& p_{N-1} & p_N\\ p_2 &  \dots & p_1 & p_N & p_{N-1} \end{array}\right)$.
\end{itemize}
This element corresponds to the sets of degrees $(0,\dots,0,1,0,0)$
and, in the particular case $N=4$, to the special choice
$(0,1,1,2)$ too.
From the discussion in Section 3 it follows that the Newton exponent of $\sigma_0$ depends
on whether $N$ is even or odd.
\begin{itemize}

\item[*] $N$ even:  $l_0=N-2$ ($\epsilon_0=e^{\frac{2\pi \imath}{N-2}}$).
It is easy to see that $p_{N-1}+p_N\in{\mathbb C}((k))$ and $\sigma_0(-p_{N-1}+p_N)=-(-p_{N-1}+p_N)$.
On the other hand since $\sigma_0$ acts on $p_j$, $j=1,2,\dots,N-2$  and
$\epsilon_0$  coincides with the previous one, we have again that
$$R_k=\sum_{j=1}^{N-2}\epsilon_0^{j\,k}p_j, \qquad k=0,1,\dots,N-3,$$
satisfy $\sigma_0(R_k)=\epsilon_0^{N-2-k}R_k$. Thus  $R$ is now given by
\begin{align}\everymath{\displaystyle}\label{RNm2beven}
\nonumber R=\sum_{k=0}^{N-3}z^kf_k(z^{N-2})&\sum_{j=1}^{N-2}\epsilon_0^{j\,k}p_j
+z^{\frac{N-2}{2}}f_{N-2}(z^{N-2})(p_{N-1}-p_{N-1})\\ \\
\nonumber +&f_{N-1}(z^{N-2})(p_{N-1}+p_N).
\end{align}

\noindent
{\bf Example} For  $N=4$
$$
R=f_0(z^2)(p_1+p_2)+zf_1(z^2)(-p_1+p_2)
 +zf_2(z^2)(-p_3+p_4)+f_3(z^2)(p_3+p_4).$$

\item[*]  $N$ odd: $l_0=2(N-2)$ ($\epsilon_0=e^{\frac{\pi \imath}{N-2}}$).
Again in this case $p_{N-1}+p_N\in{\mathbb C}((k))$ and $\sigma_0(-p_{N-1}+p_N)=-(-p_{N-1}+p_N)$.
Moreover, if we look for functions
$R_k=\sum_{j=1}^{N-2}\alpha_jp_j$ such that
$$
\sigma_0(R_k)=\epsilon_0^{2(N-2-k)}R_k,\; k=0,\dots,N-3,$$
by proceeding as in the previous cases,  we find that
 $\alpha_j=\epsilon_0^{2(N-2-k)(N-2-j)}\alpha_{N-2}=\epsilon_0^{2j\,k}\alpha_{N-2}$, so that
\begin{align}\everymath{\displaystyle}\label{RNm2bodd}
\nonumber R=\sum_{k=0}^{N-3}z^{2k}f_k&(z^{2(N-2)})\sum_{j=1}^{N-2}\epsilon_0^{2j\,k}p_j
  +z^{N-2}f_{N-2}(z^{2(N-2)})(p_N-p_{N-1})\\\\
\nonumber &+f_{N-1}(z^{2(N-2)})(p_{N-1}+p_N).
\end{align}

\noindent
{\bf Example} For $N=5$
$$\begin{array}{lll}
R&=&f_0(z^6)(p_1+p_2+p_3)+z^2f_1(z^6)(e^{\frac{2\pi i}{3}}p_1+e^{\frac{4\pi i}{3}}p_2+p_3)\\  \\
           & &+z^4f_2(z^6)(e^{\frac{4\pi i}{3}}p_1+e^{\frac{2\pi i}{3}}p_2+p_3)\\  \\
           & &+z^3f_3(z^6)(-p_4+p_5)+f_4(z^6)(p_4+p_5).
\end{array}$$

\end{itemize}

Thus, the integrable deformations (\ref{def2}), (\ref{sollenard}) are determined by the expressions
of $R$ in (\ref{RN}), (\ref{RNm1}), (\ref{RNm2a}), (\ref{RNm2beven}) or
(\ref{RNm2bodd}) depending on $\sigma_0$ and the Newton exponent $l_0$.

\vspace{0.3cm}

\noindent {\bf Acknowledgements}

\vspace{0.3cm}

The authors  wish to thank Prof. Y. Kodama  for his
interest and help during the elaboration of this work.

\end{document}